\documentstyle[epsfig,twocolumn,prl,aps]{revtex}
\begin{document}
\draft
\twocolumn[\hsize\textwidth\columnwidth\hsize\csname @twocolumnfalse\endcsname
\title{Partial Teleportation of Entanglement in the Noisy Environment}
\author{Jinhyoung Lee,$^{1,2}$
M. S. Kim,$^1$ Y. J. Park,$^2$ and S. Lee$^3$
}
\address{$^1$ School of Mathematics and Physics, 
The Queen's University of Belfast, BT7 1NN, United Kingdom
\\
$^2$ Department of Physics, Sogang University, CPO Box 1142, Seoul 
100-611, Korea
\\
$^3$ Department of Applied Mathematics, Sejong University, Seoul, Korea
}
\date{\today} \maketitle

\begin{abstract}
  Partial teleportation of entanglement is to teleport one particle of
  an entangled pair through a quantum channel.  This is conceptually
  equivalent to quantum swapping.  We consider
  the partial teleportation of entanglement in the noisy environment,
  employing the Werner-state representation of the noisy channel for
  the simplicity of calculation.  To have the insight of the many-body
  teleportation, we introduce the measure of correlation information
  and study the transfer of the correlation information and
  entanglement.  We find that the fidelity gets smaller as the
  initial-state is entangled more for a given entanglement of the
  quantum channel.  The entangled channel transfers at least some of
  the entanglement to the final state.

\end{abstract}

\vskip1pc] 

\newpage

\section{introduction}
Quantum teleportation of a single-particle state has been extensively
studied both theoretically \cite{Bennett93} and experimentally
\cite{Bouwmeester97}.  Quantum teleportation reproduces an unknown
quantum state at a remote place while the original state is destroyed
\cite{Bennett93}. The key of the quantum teleportation is the quantum
channel composed of the quantum entangled pair.  If the quantum
channel is maximally entangled, for example, by using the singlet
state, the quantum state is perfectly reproduced at the remote place
and the fidelity of the teleportation is unity.  However, in the real
world, the quantum channel lies in the noisy environment, which
degrades the entanglement of the channel.  The less is the quantum
channel entangled, the smaller is the fidelity
\cite{Popescu94,Horodecki96-1,Gisin96-2}.  It has also been found that
the fidelity of the quantum teleportation is always larger than that
of any classical communication protocol even in the noisy
environment\cite{Popescu94}.

In this paper, we are interested in partial teleportation of an
entangled state of a two spin-1/2 system.  An entangled pair of
particles are prepared by Alice who wants to teleport one of the
entangled pair to Bob as shown in Fig.~\ref{fig:ptes}.  If the quantum
channel is maximally entangled, the partial teleportation is nothing
more than entanglement swapping \cite{Zukowski93,Pan98}.  Bennett {\it
  et al.} \cite{Bennett93} argued that teleportation is a linear
operation for the perfect quantum channel and could be extended to
what is now called entanglement swapping \cite{Zukowski93}, which has
been experimentally realised \cite{Pan98}.
Entanglement swapping was considered for a more generalised
multi-particle system \cite{Bose98} and for concentration of partially
entangled states \cite{Bose99}.  In this paper, we analyse the
environmental effects on the partial teleportation of the entangled
state, considering the entanglement transfer and the fidelity.  
We define a measure of entanglement using the partial transposition.
Although this measure does not completely agree with the entropy of entanglement
for a pure state, it is useful and gives qualitative information.  As it is easily
calculated and satisfies important conditions of the measure of entanglement,
we use it to analyse the partial teleportation in this paper.
The
concept of correlation information is also introduced.  The correlation information
we define is, in general, dependent on classical and quantum correlation but
we show that it bears a simple and useful linear relation for the partial teleportation.

We assume that the initial entangled state is pure and the mixed
quantum channel is represented by the Werner state.  We show that the
calculation is much simpler as we employ the Werner-state channel
while we do not lose the generality of the treatment.
Entanglement transfer was extensively studied by Schumacher \cite{Schumacher96}
and teleportation through general channels was considered by Horodecki et al. \cite{Horodecki99}.

\section{measures of entanglement}

For a pure state $\hat{\rho}$ of a bipartite system, one can choose
the entropy $S$ of entanglement as a measure of entanglement, where
\begin{equation}
  \label{eq:eoe}
  S={\rm Tr} \hat{\rho}_a \log_2 \hat{\rho}_a = {\rm Tr} \hat{\rho}_b
  \log_2 \hat{\rho}_b
\end{equation}
where $\hat{\rho}_{a,b} = {\rm Tr}_{b,a} \hat{\rho}$ is the reduced density matrix for the subsystem $a$ or
$b$.  For a mixed state, there have been many definitions for the
measure of entanglement such as entanglement of formation
\cite{Bennett96}, quantum relative entropy, and Bures metric
\cite{Vedral97,Vedral98}.  Every measure $E(\hat{\rho})$ should
satisfy the following necessary conditions for a given density matrix
$\hat{\rho}$ \cite{Vedral97,Vedral98},
\begin{itemize}
\item[(C.1)] $E(\hat{\rho}) = 0$ if and only if $\hat{\rho}$ is
  separable.
\item[(C.2)] A local unitary transformation leaves $E(\hat{\rho})$
  invariant;
  \begin{equation}
    E(\hat{U}_1\otimes\hat{U}_2 \hat{\rho}
    \hat{U}^\dagger_1\otimes\hat{U}^\dagger_2)=E(\hat{\rho}) 
  \end{equation}
  for all unitary operators $\hat{U}_1$ and $\hat{U}_2$.
\item[(C.3)] $E(\hat{\rho})$ cannot increase under local general
  measurements (LGM), classical communications (CC), and post
  selection of subensemble (PSS),
  \begin{equation}
    \sum p_i E(\hat{\rho}_i) \le E(\hat{\rho}),
  \end{equation}
  where $p_i \hat{\rho}_i = \hat{A}_i\otimes\hat{B}_i \hat{\rho}
  \hat{A}^\dagger_i\otimes\hat{B}^\dagger_i$ with $p_i = \mbox{Tr}
  \hat{A}_i\otimes\hat{B}_i \hat{\rho}
  \hat{A}^\dagger_i\otimes\hat{B}^\dagger_i$; two set of LGM operators
  $\{\hat{A}_i\}$ and $\{\hat{B}_i\}$ are classically correlated by
  CC.
\end{itemize}
The requirement of the condition (C.1) is clear since a separable
state is just classically correlated and should be independent of the
entanglement. Since a local unitary operation is performed only
locally, it cannot affect any entanglement, required by the condition
(C.2). The condition (C.3) is related with the purification procedure
which selects a subensemble of maximally-quantum-correlated pairs
among an impure ensemble \cite{Bennett96}. The purification procedure
can distill maximally entangled states such that $E(\hat{\rho}_i) \ge
E(\hat{\rho})$ for a certain route $i$, but the average entanglement
cannot increase over the whole ensemble since quantum nonlocal
operations are not introduced, represented in the condition (C.3).

For a two spin-1/2 system, we define the measure of entanglement in
terms of the negative eigenvalues of the partial transposition of the
state. Consider a density matrix $\hat{\rho}$ for a two spin-1/2
system and its partial transposition $\hat{\sigma}= \hat{\rho}^{T_2}$.
The density matrix $\hat{\rho}$ is inseparable if and only if
$\hat{\sigma}$ has any negative
eigenvalues\cite{Horodecki96-1,Peres96-1}. The measure of entanglement
$E(\hat{\rho})$ is then defined as
\begin{equation}
  \label{eq:moet}
  E(\hat{\rho}) = - 2\sum_i \lambda^-_i
\end{equation}
where $\lambda^-_i$ are the negative eigenvalues of $\hat{\sigma}$ and
the factor 2 is introduced to have $0 \le E(\hat{\rho}) \le 1$.  In
Appendix, we prove that the entanglement measure
(\ref{eq:moet}) satisfies the above necessary conditions.  

\begin{figure}[htbp]
  \begin{center}
    \leavevmode 
    \includegraphics[height=6cm,width=8cm]{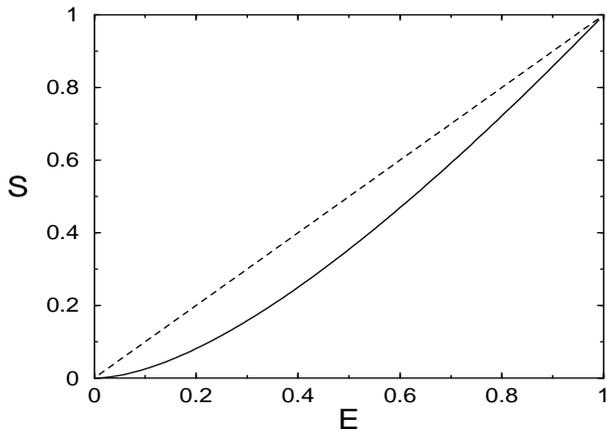}
    \caption{The entropy of entanglement $S$ (solid line) in terms of
      the measure of entanglement $E$ for a pure spin-1/2 entangled state. The
      dashed line is an eye-guidance for a linear curve. The entropy of
      entanglement $S$ is a monotonously
      increasing function of the measure of entanglement $E$ for a
      pure state because the first derivative of $S$ is positive
      everywhere from 0 to 1 of $E$.}
    \label{fig:eeoe} 
  \end{center} 
\end{figure}

In fact, there is the fourth condition which a measure of entanglement
has to satisfy:
\begin{itemize}
\item[(C.4)] 
For pure states, the measure of entanglement reduces to the entropy of
entanglement.
\end{itemize}
We note that for a pure entangled state the entanglement measure
(\ref{eq:moet}) is not reduced to the entropy of entanglement $S$ but
is a monotonously increasing function of $S$ as shown in
Fig.~\ref{fig:eeoe}.  Vedral and Plenio \cite{Vedral98} showed
that  the Bures metric satisfies the condition (C.1)-(C.3)
but is smaller than the entropy of entanglement for pure states.
They then wrote that measures which do not
satisfy condition (C.4) can nevertheless contain useful information on
entanglement.  The Schmidt norm is another example of the measure of
entanglement which does not satisfy condition (C.4) \cite{Witte99}.
The entanglement measure (\ref{eq:moet}) can qualitatively give
information on the entanglement of a given state as it satisfies
conditions (C.1)-(C.3).  Because of convenience in calculation, we use
$E(\hat{\rho})$ in Eq. (\ref{eq:moet}) as the measure of entanglement
in this paper.

\section{partial teleportation of entanglement}

\begin{figure}[htbp]
  \begin{center}
    \leavevmode
    \includegraphics*[angle=90,height=8cm,width=10cm]{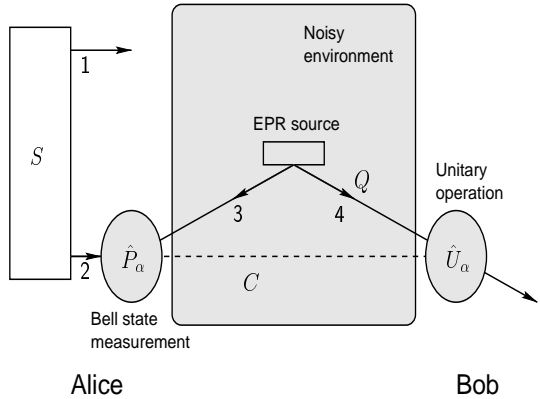}
    \caption{Schematic drawing for the partial teleportation. An
      unknown spin-1/2 two-body quantum entangled state is generated
      by Alice's source $S$. The quantum channel is produced and Alice
      and Bob share the correlated pair. Alice performs the Bell
      measurement on the particles 2 and 4 and sends the result to Bob
      through the classical channel.  Bob unitarily transforms the
      particle 4 to complete the partial teleportation.  We are
      interested in the entanglement and closeness
      of the state of particles 1 and 4 to the initial state of
      particles 1 and 2.}
  \label{fig:ptes}
  \end{center}
\end{figure}

We consider the partial teleportation of entanglement as shown in
Fig.~\ref{fig:ptes}, where Alice teleports one particle of her
entangled pair to Bob.  Alice and Bob share an ancillary pair of an
entangled state.  Alice performs the Bell-state measurement on one of
her original entangled pair and her part of the ancillary pair.  Upon
receiving Alice's measurement result through a classical channel, Bob
unitary rotates his part of the ancillary pair based on it.  If the
ancillary pair is perfectly entangled, Bob's particle and Alice's
unmeasured particle become entangled as the Alice's original entangled
pair.  The basic idea of the partial teleportation is similar to the
single-particle teleportation or entanglement swapping but our
interest here does not stop at a simple result of the teleportation of
a particle.  In this paper, the quantum channel is represented by a
mixed entangled state due to the influence from the environment.  We
are interested in how the entanglement of the teleported state is
affected by such the imperfect quantum channel by studying the
channel-dependent fidelity, information transfer and entanglement
transfer.  We assume for the simplicity that Alice's initial state is
pure and the quantum channel is represented by a Werner state
\cite{Werner89}.

An initial pure state for entangled two particles 1 and 2 is given in
the Hilbert-Schmidt space by
\begin{equation}
  \label{eq:rhoi}
  \hat{\rho}^{12} = \frac{1}{4}\left(\hat{1}\otimes \hat{1} +
    \vec{a}_0\cdot\vec{\sigma}\otimes \hat{1} + \hat{1} \otimes
    \vec{b}_0\cdot\vec{\sigma} + \sum_{nm} c_0^{nm}
    \hat{\sigma}_n\otimes\hat{\sigma}_m \right)
\end{equation}
where $\vec{a}_0$ and $\vec{b}_0$ are real vectors and $c_0^{nm}$ is
an element of the real matrix $C_0$.  The
initial pure state $\hat{\rho}_0$ in Eq.~(\ref{eq:rhoi}) satisfies the pure
state condition $\hat{\rho}^2=\hat{\rho}$.  
We can also consider a general
representation of the initial pure state (\ref{eq:rhoi}), with help of
the seed state $\hat{\rho}^{12}_s$ defined as follows
\begin{equation}
  \label{eq:rhoi2}
  \hat{\rho}^{12} = (\hat{U}^1 \otimes \hat{U}^2) \hat{\rho}_s^{12}
  (\hat{U}^1 \otimes \hat{U}^2)^\dagger 
\end{equation}
where $\hat{U}^1$ and $\hat{U}^2$ are local unitary operators acting
respectively on the particles 1 and 2.  The density operator for the
seed state is
\begin{equation}
  \label{eq:sdm}
  \hat{\rho}_s^{12} = \frac{1}{4}\left(\hat{1}\otimes \hat{1} +
    a_0\hat{\sigma}_z\otimes \hat{1} + 
    \hat{1} \otimes a_0\hat{\sigma}_z + \sum_n c_n
    \hat{\sigma}_n\otimes\hat{\sigma}_n \right) 
\end{equation}
where $a_0$ is a positive real number and $\vec{c}=(c_0,-c_0,1)$ a
real vector, constrained by $a_0^2+c_0^2=1$.  The vector $\vec{c}$
describes the quantum correlation of the pure state
$\hat{\rho}_s^{12}$, yielding the relation $E_0=|c_0|$, where $E_0$ is
the measure of entanglement for $\hat{\rho}_s^{12}$.  The state has no
entanglement, i.e., $E_0=0$, if and only if $c_0=0$.  Now, the state
$\hat{\rho}_s^{12}$ is characterised by one parameter $c_0$ or
equivalently by its entanglement $E_0$.  By the definition of the
measure of entanglement, the initial state of the density operator
$\hat{\rho}^{12}$ and the state of the seed density operator
$\hat{\rho}_s^{12}$ have the same measure of entanglement, $E_0$.  It
is thus clear that the initial state $\hat{\rho}^{12}$ is fully
determined by its measure of entanglement $E_0$ and the local unitary
operations $\hat{U}^1$ and $\hat{U}^2$.

We take the Werner state for the quantum channel.  In fact, any mixed
state can be made a Werner state by random local $SU(2)\otimes SU(2)$
operations \cite{Bennett96,Werner89} so that we do not lose the
generality by taking the Werner state in describing the mixed channel.
The Werner state $\hat{w}^{34}$ of the ancillary particles 3 and 4 is
\begin{equation}
  \label{eq:qc}
  \hat{w}^{34} = \frac{1}{4}\left(\hat{1}\otimes \hat{1} +
  \sum_{nm} c_w^{nm} \hat{\sigma}_n\otimes \hat{\sigma}_n\right).
\end{equation}
where $c_{w}^{nm}$ is an element of 
the real matrix $C_w = (2\Phi+1)/3~\cdot~{\rm diag}(-1,-1,-1)$. 
The Werner
state becomes the singlet state when $\Phi=1$. The parameter $\Phi$ is
related to the entanglement $E_w$ of the Werner state $\hat{w}^{34}$. 
It is straightforward to show that $E_w = \mbox{max}(0, \Phi)$.

The Bell-state measurement by Alice is represented by a family of
projectors
\begin{equation}
  \label{eq:bsmp}
  \hat{P}_{\alpha}^{23} = |\Psi_{\alpha}^{23}\rangle 
\langle\Psi_{\alpha}^{23}| =
  \frac{1}{4} \left(I\otimes I + 
  \sum_{nm} p_\alpha^{nm} \sigma_n \otimes \sigma_m\right) 
\end{equation}
where $|\Psi_\alpha^{23}\rangle$ are the four possible Bell states and
$p_\alpha^{nm}$ is an element of the real matrix $P_\alpha$;
$P_0=\mbox{diag}(-1,-1,-1)$, $P_1=\mbox{diag}(-1,1,1)$,
$P_2=\mbox{diag}(1,-1,1)$, $P_3=\mbox{diag}(1,1,-1)$.  The Bell
measurement is performed on the particle 2 of the initial entangled
pair and the particle 3 of the Werner state (see Fig.~\ref{fig:ptes}).

The quantum teleportation utilises both the classical and quantum
channels.  Upon receiving the two-bit classical message on the
Bell-state measurement through the classical channel, Bob performs the
unitary transformation on the particle 4 accordingly.  If the quantum
channel is in the spin singlet state, the teleportation can be
perfectly completed by one of the following four possible unitary
operators: $\hat{1}$, $\hat{\sigma}_x$, $\hat{\sigma}_y$, and
$\hat{\sigma}_z$.  However, for the mixed channel, it is difficult for
Bob to decide which unitary transformation to get the final state
$\hat{\rho}^{14}$ maximally close to the initial state
$\hat{\rho}^{12}$.  Let us consider how to determine the right unitary
transformation when the channel is in the Werner state.

Suppose Bob receives a two-bit message through the classical channel
saying that Alice's measurement was $|\Psi^{23}_\alpha\rangle$.  Bob
then applies the unitary transformation $\hat{U}^4_\alpha$ on the
particle 4, then the state $\hat{\rho}^{14}_\alpha$ of the two
particles 1 and 4 becomes
\begin{equation}
  \label{eq:bsos}
  \hat{\rho}^{14}_\alpha = \frac{1}{p_\alpha} {\rm Tr}_{2,3} \left 
    [ \hat{P}^{23}_\alpha \otimes \hat{U}^4_\alpha \left
    ( \hat{\rho}^{12}_0 \otimes \hat{w}^{34} \right)
    \left. \hat{P}^{23}_\alpha \otimes \hat{U}^4_\alpha
    \right.^\dagger \right]
\end{equation}
where $p_\alpha$ is the probability of $|\Psi^{23}_\alpha\rangle$ to
be the result of Alice's measurement.  Eq.~(\ref{eq:bsos}) can be
written in the Hilbert-Schmidt space as
\begin{equation}
  \label{eq:rhor}
  \hat{\rho}_\alpha^{14} = \frac{1}{4}\left(\hat{1}\otimes \hat{1} +
    \vec{a}_\alpha\cdot\vec{\sigma}\otimes \hat{1} + \hat{1}\otimes 
    \vec{b}_\alpha\cdot\vec{\sigma} +
    \sum_{nm}\tilde{c}^{nm}_\alpha\hat{\sigma}_n\otimes\hat{\sigma}_m\right).
\end{equation}
with the parameters 
\begin{eqnarray}
  \label{eq:rhorpm}
 \vec{a}_\alpha &=& \vec{a}_0,\nonumber \\ 
  \vec{b}_\alpha &=& \frac{2\Phi+1}{3} O^T_\alpha P_\alpha
  \vec{b}_0, \nonumber \\
  C_\alpha &=& \frac{2\Phi+1}{3} O^T_\alpha
  P_\alpha C_0.
\end{eqnarray}
Here we have the rotation matrix $O_\alpha$ in the Bloch space for a
single particle state, obtained from the unitary operator
$\hat{U}_\alpha$ \cite{Horodecki96-1}:
\begin{equation}
  \label{eq:rmfum}
  \hat{U}_\alpha \vec{a} \cdot \vec{\sigma} \hat{U}_\alpha^\dagger =
    \left(O_\alpha^T \vec{a}\right) \cdot \vec{\sigma}.
\end{equation}

The fidelity $F$ measures how close the final state $\hat{\rho}^{14}$
is to the initial state $\hat{\rho}^{12}$; $F=\sum_\alpha p_\alpha
{\rm Tr}\hat{\rho}^{12} \hat{\rho}_\alpha^{14}$.  If the teleportation
is perfect, the final state is the same as the initial state so that
the fidelity is 1.  By substituting $\hat{\rho}^{12}$ in
(\ref{eq:rhoi}) and $\hat{\rho}_\alpha^{14}$ in (\ref{eq:rhorpm}) into
the definition of the fidelity, we find the fidelity for the
Werner-state channel
\begin{eqnarray}
  \label{eq:fpet}
 F &=& \frac{1}{4} \left[ 1 + |\vec{a}_0|^2 + \frac{2\Phi+1}{3}
  \vec{b}_0 \cdot \left( -\frac{1}{4}\sum_\alpha O^T_\alpha P_\alpha
  \vec{b}_0 \right) \right. \nonumber \\ 
  & & + \left.  \frac{2\Phi+1}{3} \mbox{Tr}\left(-\frac{1}{4}
    \sum_{\alpha} O^T_\alpha P_\alpha C_0^T C_0 \right) \right].
\end{eqnarray}

The task is now to find Bob's unitary operations $\hat{U}_\alpha$ to
maximise fidelity (\ref{eq:fpet}).  For a general mixed channel, the
fidelity is a function of the initial and channel states as well as
Bob's unitary operation.  However, the basic assumption of the quantum
teleportation is that the initial state is unknown.  In order to
examine the faithfulness of quantum teleportation, we need to average
the fidelity over the Hilbert space where the initial state lies in.
The unitary operations should be determined to maximise the average
fidelity \cite{Horodecki96-1}.

For the Werner-state channel the fidelity has been calculated as in
Eq.~(\ref{eq:fpet}), where the measurement dependence is found in the
terms including $-\sum_\alpha O_\alpha^T P_\alpha$.  It is clear that
$-P_\alpha$ in Eq.~(\ref{eq:bsmp}) is a rotation matrix thus
$|O^T_\alpha P_\alpha|\le 1$.  Choosing $O_\alpha=-P_\alpha$ to
maximise the fidelity (\ref{eq:fpet}), we find that the corresponding
unitary operators are the same as in the singlet-state channel
discussed above.  This choice enables Bob to produce the
measurement-independent final state $\hat{\rho}^{14}_\alpha=
\hat{\rho}^{14}$.

The fidelity for the Werner-state channel is then given by
\begin{equation}
  \label{eq:mfpet}
  F = \frac{1}{4} \left( 1 + |\vec{a}_0|^2 + \frac{2\Phi+1}{3}
  |\vec{b}_0|^2 + \frac{2\Phi+1}{3} \mbox{Tr} C_0^T C_0 \right). 
\end{equation}
It is seen that the fidelity is invariant for the local unitary
transformations on the initial pure state.  Noting $|\vec{a}_0|$,
$|\vec{b}_0|$ and $\mbox{Tr}C^T_0C_0$ are uniquely determined by the
entanglement $E_0$ for the initial state, the fidelity can be finally
written as
\begin{equation}
\label{eq:fidelity-ent}
  F = \frac{E_w+2}{3} + \frac{E_w-1}{6} E_0^2.
\end{equation}
It is clear that the fidelity (\ref{eq:fidelity-ent}) depends on the
initial-state entanglement $E_0$ and channel entanglement $E_w$.

The large entanglement in the channel enhances the fidelity and the
maximally entangled channel gives the unit fidelity independent from
the initial-state entanglement $E_0$. When Alice's initial state is
disentangled, {\it i.e.} $E_0=0$, it can be written as a direct
product of two individual states and the partial entanglement
teleportation becomes equivalent to a single-particle case. In this
case, the fidelity is $F= (E_w+2)/3$, equal to that for the
single-particle teleportation \cite{Horodecki96-1}. It has an upper
bound of $2/3$ producible by classical communication of $E_w=0$
\cite{Popescu94}.  For the entangled initial state with $E_0 \ne 0$,
the large initial-state entanglement $E_0$ reduces monotonously the
fidelity (\ref{eq:fidelity-ent}) for a given channel entanglement
$E_w<1$ because the sign of the second term is negative in
Eq.~(\ref{eq:fidelity-ent}).  This implies that the initial
entanglement has fragile nature to teleport. In other words, the
entanglement is destroyed easily by the environment.

To examine the loss of the initial entanglement, we consider the
entanglement transfer by teleportation.  We are interested in how much
the initial-state entanglement is transferred to the final state.  The
measure of entanglement for the final state is calculated as
\begin{equation}
  \label{eq:fers}
  E(\hat{\rho}^{14}) = \frac{1}{3}
  \left[\sqrt{(1-E_w)^2+3E_w(2+E_w)E_0^2} - (1-E_w)\right]. 
\end{equation}
As $\partial E/\partial E_w \ge 0$, large entanglement of the channel
enhances the entanglement transfer to the final state from the initial
one. For a given entanglement of the channel, the entanglement of the
final state increases as the initial-state entanglement gets larger.
The entanglement of the final state is nonzero as far as $E_w \ne 0$
and $E_0 \ne 0$, which shows that the entangled channel transfers at
least some of the initial entanglement to the final state.

\section{correlation information}

Brukner and Zeilinger have recently derived a measure of information
for a quantum state \cite{Brukner99}. The information measure is
invariant for a choice of a complete set $\{\hat{A}_1,\cdot\cdot\cdot,
\hat{A}_m\}$ of complementary observations and is conserved as far as
there is no information exchange between the system and the
environment \cite{Brukner99}. We employ the measure of information to
study the quantum information transfer in the partial teleportation of
entanglement.

Suppose an experimental arrangement for a measurement by observable
$\hat{A}_j$ which has $n$ possible outcomes with $n$ dimensional
probability vector $\vec{p}=(p_1, ..., p_i, ..., p_n)$ for a given
system. The system is supposed to have maximum $k$-bits of information
such that $n=2^k$. The measure of information $I_j$ for the observable
$\hat{A}_j$ is defined as
\begin{equation}
  \label{eq:imqi}
  I_j = {\cal N} \sum_{i=1}^n \left(p_i - \frac{1}{n}\right)^2.
\end{equation}
where the normalisation constant ${\cal N}=2^kk/(2^k-1)$. $I_j$
results in $k$ bits of information if one $p_i=1$ and 0 bits of
information if all $p_i$ are equal.  For a complete set of $m$
mutually complementary observables the measure of information is
defined as the sum of the measures of information over the complete
set
\begin{equation}
  \label{eq:tmqi}
  I(\hat{\rho}) = \sum_{j=1}^m I_j.
\end{equation}
for the quantum state $\hat{\rho}$. A single spin-1/2 system, for
example, is represented by the measure of information $I(\hat{\rho}) =
2 {\rm Tr} \hat{\rho}^2 - 1$.

In the following, we define the measure of correlation information
based on the measure of information introduced by Brukner and
Zeilinger.  Let us consider the measure of information for a composite
system of two particles which can be decomposed into three parts.
Each particle has its own information corresponding to its reduced
density matrix, which we call the {\it individual information}.  The
two particles can also have the correlation information which depends
on how much they are correlated.

The measure of total information for a density matrix $\hat{\rho}$ of
the two spin-1/2 particles is \cite{Brukner99}
\begin{equation}
  \label{eq:tqim}
  I(\hat{\rho}) = \frac{2}{3} \left(4{\rm Tr}\hat{\rho}^2 - 1\right).
\end{equation} 
The measures of individual information $I_a(\hat{\rho})$ and
$I_b(\hat{\rho})$ for the particles $a$ and $b$ are
\begin{eqnarray}
  \label{eq:oqia}
  I_a(\hat{\rho}) &=& 2 {\rm Tr}_a \left(\hat{\rho}_a\right)^2 - 1 \\ 
  \label{eq:oqib}
  I_b(\hat{\rho}) &=& 2 {\rm Tr}_b \left(\hat{\rho}_b\right)^2 - 1
\end{eqnarray}
where $\hat{\rho}_{a,b} = {\rm Tr}_{b,a} \hat{\rho}$ are reduced
density matrices for the particles $a$ and $b$.  If the total density
matrix $\hat{\rho}$ is represented by $\hat{\rho}=
\hat{\rho}_a\otimes\hat{\rho}_b$, the total system is completely
separable and we know that there is no correlation information.  We
thus define the measure of correlation information as
\begin{eqnarray}
  \label{eq:jqim}
I_c(\hat{\rho}) = I(\hat{\rho}) - I(\hat{\rho}_a\otimes\hat{\rho}_b)
  = I(\hat{\rho}) - \frac{2}{3}[ I_a(\hat{\rho})
    + I_b(\hat{\rho}) \nonumber \\ + \mbox{}I_a(\hat{\rho})I_b(\hat{\rho})]. 
\end{eqnarray}
The measures of individual and correlation information are invariant
for any particular choice of the complete set of complementary
observables.

If there is no correlation between the two particles, the measure of
total information is a mere sum of the measures of individual
information.  On the other hand, the total information is imposed only
on the correlation information, $I=I_c$, if there is no individual
information, $I_a=I_b=0$. For a two spin-1/2 system, the maximally
entangled states have only the correlation information.

%The entanglement teleportation transfers the correlation information as
%well as the individual information of the initial state. Because the quantum
%channel is coupled with the environment, some of the information
%would leak to the environment through the quantum channel. The
%information transfer can provide us with good insight on the entanglement
%teleportation. The correlation-information transfer is especially related to
%the entanglement transfer. 

Note that the correlation information is not the same as the measure of
entanglement.  Only when a pure state is considered the measure of the
correlation is directly related to the measure of entanglement.  For a mixed
state, the correlation information also includes the information due
to classical correlation.  For example, the state of the density
operator $\hat{\rho}_{cc}=1/4(\hat{1}\otimes \hat{1}- \sigma_z\otimes
\sigma_z)$ is not quantum-mechanically entangled but classically
correlated with the correlated information $I_c\neq 0$ \cite{cc}.

The total information $I(\hat{\rho}^{14})$ in the final state
(\ref{eq:rhor}) is obtained using its definition (\ref{eq:tqim}):
\begin{equation}
  \label{eq:tior}
  I(\hat{\rho}^{14}) = \frac{2}{3} \left[ 1 + 2
    \left(\frac{2E_w+1}{3}\right)^2 + \left\{ 
    \left(\frac{2E_w+1}{3}\right)^2 - 1 \right\}E_0^2 \right]
\end{equation}
which depends on the initial-state and the channel entanglement.  We
can easily find that $0\le I(\hat{\rho}^{14}) \le 2$ from the range of
the entanglement measure $0\le E_w,~E_0\le 1$.  The measure of
information for the initial state is 2 as it is a pure two spin-1/2
system.  The final state can have at best the same amount of
information as the initial state because the noisy environment acts
only to dissipate the information.  The total information is better
preserved for the larger channel entanglement $E_w$.  The total
information is lost more easily for the larger initial entanglement as
the sign of the coefficient for $E_0^2$ is negative. This is
consistent with the discussions for the fidelity.

We evaluate the measures of individual $I_1$, $I_4$ and correlation
$I_c$ information for the final state given by
\begin{eqnarray}
  \label{eq:aoir}
  I_1(\hat{\rho}^{14}) &=& I_1^0 \\
  \label{eq:boir}
  I_4(\hat{\rho}^{14}) &=& \left( \frac{2E_w+1}{3} \right)^2 I_2^0 \\
  \label{eq:joir}
  I_c(\hat{\rho}^{14}) &=& \left( \frac{2E_w+1}{3} \right)^2 I_c^0
\end{eqnarray}
where $I_1^0=1-E_0^2$, $I_2^0=1-E_0^2$, and $I_c^0=2(4-E_0^2)E_0^2/3$
are the measures of individual and correlation information for the
initial state $\hat{\rho}^{12}$.  Because there has been no action on
Alice's particle 1, its individual information remains unchanged with
$I_1=I_1^0$. On the other hand, the individual information for the
particle 2 is not fully transferred to the particle 4 and the
correlation information is decreased.  The coefficient for the
decrease of the information is the same for $I_4$ and $I_c$ but we
must remember that the maximum measure of correlation information is 2
while that of individual information is 1, which shows that the
correlation information can be lost more easily.

Because the final state is a mixed state, its correlation information
$I_c$ describes in general both the quantum and classical
correlations.  For an entangled initial state of $E_0 \ne 0$ we
consider two cases: $E_w = 0$ and $E_w \ne 0$. If $E_w=0$, the final
state is classically correlated because $I_c \ne 0$ in
Eq.~(\ref{eq:joir}) whereas $E=E(\hat{\rho}^{14})=0$ in
Eq.~(\ref{eq:fers}).  On the other hand, if $E_w \ne 0$, the
correlation information $I_c$ can be written in terms of the final
entanglement $E$:
\begin{eqnarray}
  I_c = 2 \left( \frac{2E_w+1}{3} \right)^2 \left( 4- 3
  \frac{E+(1-E_w)}{E_w(2+E_w)} E\right) \nonumber \\ \times \frac{E+(1-E_w)}{E_w(2+E_w)} E.
\end{eqnarray}
The correlation information $I_c = 0$ if and
only if $E=0$ for partial teleportation via
the Werner channel. This shows that for $E_w \ne 0$ the correlation
information of the final state is only due to the quantum correlation.

\section{Remarks}

The partial teleportation of entanglement has been considered in the
noisy environment. The measures of individual and correlation
information have been extensively studied for the two spin-1/2 system.
As the Werner-state is employed for the quantum channel, the
calculation of the fidelity, the information transfer and the
entanglement transfer became extremely simple while we do not lose the
generality to consider the noisy environment.  The larger the
initial-state entanglement is, the worse the fidelity becomes for any
imperfect quantum channel.  We, however, find more entanglement in the
final state with the larger initial-state entanglement.  The entangled
channel transfers at least some of the entanglement to the final
state.

\acknowledgements
JL thanks Mr.\v{C}. Bruckner for discussions at the Entanglement and
Decoherence Workshop in Garda.
This work was supported by the Brain Korea 21 grant (D-0055) of the Korean
Ministry of Education.

\appendix

\section*{proof of the measure of entanglement Eq.~(4)}

Consider a density matrix $\hat{\rho}$ for a two spin-1/2 system and
the partial transposition $\hat{\sigma}= \hat{\rho}^{T_2}$
\cite{Note1}. The density matrix $\hat{\rho}$ is inseparable if and
only if $\hat{\sigma}$ has any negative eigenvalues
\cite{Horodecki96-1,Peres96-1}. The measure of entanglement ${\cal
  E}(\hat{\rho})$ is defined as $2\sum_i (-\lambda^-_i)$ with the
negative eigenvalues $\lambda^-_i$ of $\hat{\sigma}$. We will show
that ${\cal E}(\hat{\rho})$ satisfies the three conditions
(C.1)-(C.3).

Let $\hat{d}(\hat{\rho})$ be a diagonal matrix of the partial
transposition $\hat{\sigma}$ such that, for some unitary operator $U$,
\begin{eqnarray}
  \hat{d}(\hat{\rho}) &=& \left(U \hat{\sigma} U^\dagger \right)
  \nonumber \\ &=& \mbox{diag}(\lambda_1, \lambda_2, \lambda_3,
  \lambda_4)
\end{eqnarray}
where $\mbox{diag}(\{\lambda_i\})$ represents a diagonal matrix with
its diagonal elements $\lambda_i$ and thus $\lambda_i$ are eigenvalues
of $\hat{\sigma}$. For the given $\hat{d}(\hat{\rho})$, the density
matrix space is decomposed into two subspaces; one is expanded by
eigenvectors of the semi-positive (positive or zero) eigenvalues and
the other of the negative eigenvalues of $\hat{d}$.  The identity
operator $\hat{1}$ is then the sum of two projectors $\hat{P}_+$ and
$\hat{P}_-$ such that
\begin{equation}
  \hat{1} = \hat{P}_+ + \hat{P}_-
\end{equation}
where $\hat{P}_+$ ($\hat{P}_-$) projects the density matrix space onto
the semi-positive (negative) eigenvalue subspace. Any hermitian matrix
$\hat{H}$ is decomposed into
\begin{equation}
  \hat{H} = \hat{P}_+ \hat{H} \hat{P}_+ + \hat{P}_+ \hat{H} \hat{P}_-
  + \hat{P}_- \hat{H} \hat{P}_+ + \hat{P}_- \hat{H} \hat{P}_-
\end{equation}
and thus
\begin{equation}
  \hat{d}(\hat{\rho}) = \hat{d}_+(\hat{\rho}) + \hat{d}_-(\hat{\rho})
\end{equation}
where $\hat{d}_+ \equiv \hat{P}_+ \hat{d} \hat{P}_+$ and $\hat{d}_-
\equiv \hat{P}_- \hat{d} \hat{P}_-$. Note that $\hat{P}_+ \hat{d}
\hat{P}_- = \hat{P}_- \hat{d} \hat{P}_+ = 0$ since $\hat{d}$ is a
diagonal matrix. Now, the measure of entanglement ${\cal
  E}(\hat{\rho})$ is defined as twice the absolute value of the trace
on the negative diagonal matrix $\hat{d}_-(\hat{\rho})$, given by
\begin{equation}
  \label{meavtnd}
  {\cal E}(\hat{\rho}) \equiv - 2 \mbox{Tr} \left[
    \hat{d}_-(\hat{\rho}) \right] = 2 \sum_\beta (-\lambda^-_\beta)
\end{equation}
where $\lambda^-_\beta$ is negative eigenvalue of $\hat{\sigma}$ and
the factor 2 is introduced to be $0 \le {\cal E}(\hat{\rho}) \le 1$.

Now we consider that $E(\hat{\rho})$ in Eq.~(\ref{eq:moet}) satisfies
the necessary conditions (C.1)-(C.3). If $\hat{\rho}$ is separable,
$\hat{\sigma}$ has no negative eigenvalues and the converse statement
also holds, satisfying the condition (C.1). A local unitary
transformation leads to new density matrix $\hat{\rho}^\prime =
\hat{U}_1 \otimes \hat{U}_2 \hat{\rho} \hat{U}^\dagger_1 \otimes
\hat{U}^\dagger_2$ and its partial transposition $\hat{\sigma}^\prime
= \hat{U}_1 \otimes \hat{U}^*_2 \hat{\sigma} \hat{U}^\dagger_1 \otimes
(\hat{U}^*_2)^\dagger$. Note that $\hat{U}_1 \otimes \hat{U}^*_2$ is a
unitary operator such that $\hat{U}_1 \otimes \hat{U}^*_2
\hat{U}^\dagger_1 \otimes (\hat{U}^*_2)^\dagger = \hat{1}$. Since the
eigenvalues are independent of the unitary transformation, the
condition (C.2) is satisfied with $E(\hat{\rho}^\prime) =
E(\hat{\rho})$.

To consider the final condition (C.3), we introduce the LGM+CC(+PSS)
which maps the density matrix $\hat{\rho}$ into $\hat{\rho}'$, defined
by
\begin{equation}
  \hat{\rho}' = \sum_i \hat{V}_i \hat{\rho} \hat{V}^\dagger_i
\end{equation}
where the classically correlated operator $\hat{V}_i = \hat{A}_i
\otimes \hat{B}_i$ satisfies the complete relation as $\sum_i
(\hat{V}_i)^\dagger \hat{V}_i = \hat{1}$. Let $p_i \hat{\rho}_i =
\hat{V}_i \hat{\rho} \hat{V}^\dagger_i$ with $p_i = \mbox{Tr}\hat{V}_i
\hat{\rho} \hat{V}^\dagger_i$ and $\hat{\sigma}_i =
\hat{\rho}^{T_2}_i$. Since $\hat{V}_i$ is a local operator,
$\hat{\sigma}_i$ is represented in terms of $\hat{\sigma}$, namely,
\begin{equation}
  p_i \hat{\sigma}_i = \hat{V}^\prime_i \hat{\sigma}
  \left.\hat{V}^\prime_i\right.^\dagger
\end{equation}
where $\hat{V}^\prime_i = \hat{A}_i \otimes \hat{B}^*_i$ is an LGM+CC
operator with a completeness relation $\sum_i
\left.\hat{V}^\prime_i\right.^\dagger \hat{V}^\prime_i = \hat{1}$.

Suppose $\hat{d}^i$ are diagonal matrices of $\hat{\sigma}_i$ with
some unitary operator $\hat{U}_i$ and $\hat{d}$ of $\hat{\sigma}$ with
$\hat{U}$. The diagonal matrix $\hat{d}^i$ can be written as
\begin{eqnarray}
  p_i \hat{d}^i &=& \left(\hat{U}_i \hat{V}^\prime_i \hat{U}^\dagger\right)
  \left(\hat{U} \hat{\sigma} \hat{U}^\dagger\right) \left(\hat{U}
  \left.\hat{V}^\prime_i\right.^\dagger \hat{U}_i^\dagger\right)
  \nonumber \\ &=& 
  \hat{W}_i \hat{d} \hat{W}^\dagger_i
\end{eqnarray}
where $\hat{W}_i = \hat{U}_i \hat{V}^\prime_i \hat{U}_0^\dagger$ is
also a LGM+CC operator. For the given $\hat{d}^i$, two projectors
$\hat{P}^i_-$ and $\hat{P}^i_+$ are defined to project the density
matrix space onto semi-positive and negative eigenvalue subspaces of
$\hat{d}^i$. The measure of entanglement $E(\hat{\rho}_i)$ on the
subensemble $\hat{\rho}_i$ is given by
\begin{eqnarray}
  p_i E(\hat{\rho}_i) &=& - 2 p_i \mbox{Tr} \left[
    \hat{d}^i_-(\hat{\rho}_i) \right] \nonumber \\ &=& - 2 \mbox{Tr}
  \left(\hat{P}^i_- \hat{W}_i \hat{d} \hat{W}^\dagger_i \hat{P}^i_-
  \right).
\end{eqnarray}
We represent $\hat{d} = \sum_j \lambda_j |\psi_j \rangle \langle
\psi_j| = \sum_\alpha \lambda^+_\alpha |\psi_\alpha \rangle \langle
\psi_\alpha| + \sum_\beta \lambda^-_\beta |\psi_\beta \rangle \langle
\psi_\beta|$ where the sum is decomposed into two sums of semi-positive and
negative eigenvalues. The whole-ensemble average of the measures of
entanglement is given by
\begin{eqnarray}
  \label{eq:eae}
  \sum_i p_i E(\hat{\rho}_i) &=& 2 \sum_{ij} (-\lambda_j) \langle
  \psi_j| \hat{W}^\dagger_i \hat{P}^i_- \hat{P}^i_- \hat{W}_i |\psi_j
  \rangle \nonumber \\ &=& 2 \sum_{i\alpha} (-\lambda^+_\alpha)
  \langle \psi_\alpha| \hat{W}^\dagger_i \hat{P}^i_- \hat{P}^i_-
  \hat{W}_i |\psi_\alpha \rangle \nonumber \\ & \! & + 2 \sum_{i\beta} (-\lambda^-_\beta)
  \langle \psi_\beta| \hat{W}^\dagger_i \hat{P}^i_- \hat{P}^i_-
  \hat{W}_i |\psi_\beta \rangle \nonumber \\ 
\end{eqnarray}
where we separate the sum into the sums of semi-positive eigenvalues
$\lambda^+_\alpha$ and negative eigenvalues $\lambda^-_\beta$ of
$\hat{d}$.  The inequality, $\langle \psi_j| \hat{W}^\dagger_i
\hat{P}^i_- \hat{P}^i_- \hat{W}_i |\psi_j \rangle \ge 0$, and the sign
of eigenvalues make the first term negative and the second term
positive in Eq.~(\ref{eq:eae}). Eq.~(\ref{eq:eae}) results in the
following inequality
\begin{equation}
  \sum_i p_i E(\hat{\rho}_i) \le 2 \sum_{i\beta}
  (-\lambda^-_\beta) \langle \psi_\beta| \hat{W}^\dagger_i \hat{P}^i_-
  \hat{P}^i_- \hat{W}_i |\psi_\beta \rangle.
\end{equation}
Since $0 \le \sum_i \langle \psi| \hat{W}^\dagger_i \hat{P}^i_-
\hat{P}^i_- \hat{W}_i |\psi \rangle \le 1$ for arbitrary wave function
$|\psi\rangle$ \cite{Note2}, we finally arrive at the inequality
\begin{eqnarray}
  \sum_i p_i E(\hat{\rho}_i) & \le & 2 \sum_{\beta}
  (-\lambda^-_\beta) \sum_i \langle \psi_\beta| \hat{W}^\dagger_i
  \hat{P}^i_- \hat{P}^i_- \hat{W}_i |\psi_\beta \rangle \nonumber \\ 
  & \le & 2 \sum_{\beta} (-\lambda^-_\beta) = E(\hat{\rho}),
\end{eqnarray}
which satisfies the condition (C.3) for an arbitrary set of LGM+CC
operators.

As an example how to calculate the measure of entanglement, consider
the Werner state $\hat{w}$ in Eq.~(\ref{eq:qc}). 
When we select the following representations,
\begin{equation}
\label{eq:tshmr}
  \begin{array}{ll}
  \hat{1} \otimes \hat{1} = \left( 
    \begin{array}{cccc}
      1 & 0 & 0 & 0 \\
      0 & 1 & 0 & 0 \\
      0 & 0 & 1 & 0 \\
      0 & 0 & 0 & 1 \\
    \end{array} \right), &
 \hat{\sigma}_x \otimes \hat{\sigma}_x = \left( 
    \begin{array}{cccc}
      0 & 0 & 0 & 1 \\
      0 & 0 & 1 & 0 \\
      0 & 1 & 0 & 0 \\
      1 & 0 & 0 & 0 \\
    \end{array} \right), \\
 \hat{\sigma}_y \otimes \hat{\sigma}_y = \left( 
    \begin{array}{cccc}
      0 & 0 & 0 &-1 \\
      0 & 0 & 1 & 0 \\
      0 & 1 & 0 & 0 \\
     -1 & 0 & 0 & 0 \\
    \end{array} \right), &
 \hat{\sigma}_z \otimes \hat{\sigma}_z = \left( 
    \begin{array}{cccc}
      1 & 0 & 0 & 0 \\
      0 &-1 & 0 & 0 \\
      0 & 0 &-1 & 0 \\
      0 & 0 & 0 & 1 \\
    \end{array} \right),
  \end{array} \nonumber
\end{equation}
the Werner state is written as
\begin{equation}
  \hat{w} = \left( 
    \begin{array}{cccc}
      \frac{1-f}{4} & 0 & 0 & 0 \\ 
      0 & \frac{1+f}{4} & -\frac{f}{2} & 0 \\ 
      0 & -\frac{f}{2} & \frac{1+f}{4} & 0 \\ 
      0 & 0 & 0 & \frac{1-f}{4} \\ 
    \end{array} \right)
\end{equation}
where $f=(2\Phi+1)/3$, and the set of eigenvalues is
$\{(1-f)/4,(1-f)/4,(1-f)/4,(1+3f)/4\}$. The positivity of density
matrix requires that $-1/3 \le f \le 1$ and thus $-1 \le \Phi \le 1$.
The partial transposition is now given by
\begin{equation}
  \label{eq:wspt}
  \hat{\sigma} = \hat{w}^{T_2} = \left( 
    \begin{array}{cccc}
      \frac{1-f}{4} & 0 & 0 & -\frac{f}{2} \\ 
      0 & \frac{1+f}{4} & 0 & 0 \\ 
      0 & 0 & \frac{1+f}{4} & 0 \\ 
      -\frac{f}{2} & 0 & 0 & \frac{1-f}{4} \\ 
    \end{array} \right)
\end{equation}
which has its eigenvalues $\{(1+f)/4,(1+f)/4,(1+f)/4,(1-3f)/4\}$.  It is
clear that three eigenvalues are positive since $(1+f)/4 \ge 0$ under
the constraint of $-1/3 \le f \le 1$. The other eigenvalue can be
negative only if $3f>1$ or equivalently $\Phi>0$.  For $3f > 1$,
$E(\hat{\rho}) \equiv 2 \sum_\beta (-\lambda^-_\beta) = (3f-1)/2$
while $E(\hat{\rho}) = 0$ for $3f \le 1$. The Werner state with $f=1$
becomes a maximally entangled singlet state and then $E(\hat{w})=1$.

\newpage

\begin{references}
\bibitem{cc} The density operator for the classically-correlated state
  $\hat{\rho}_{cc}$ is written in more conventional representation as
  $\hat{\rho}=(1/2)(|\uparrow\downarrow\rangle
  \langle\downarrow\uparrow| + |\downarrow\uparrow\rangle
  \langle\uparrow\downarrow|)$.
\end{references}
\end{document}